\documentclass[10pt]{article}
\usepackage{epsfig,enumerate,eucal}
\usepackage{amssymb,amsmath,amsthm,epsf}
\usepackage{graphicx,epic,eepic,cancel,rotating}
\usepackage{subfigure}
\usepackage{comment}
\usepackage{color}

\newcommand{\n}{\noindent}
\newcommand{\ds}{\displaystyle}

\newcommand{\vp}{\varepsilon}

\newcommand{\bb}[1]{\mathbb{#1}}

\newcommand{\ms}{\medskip}

\newcommand{\crossarrow}[1]{{\text{\raisebox{2ex}{\begin{turn}{-160}{$\hbox to 35pt{\rightarrowfill}$}\end{turn}}}\hspace{-.4in}}{#1}}

\excludecomment{omitext}

\begin{document}

\title{The analytical solution of three-layer Hele-Shaw flows with linear viscous profile in the middle layer}

\title{Analytical solution of a non-standard eigenvalue problem arising in models of enhanced oil recovery}
\title{Analytical solution of a non-standard eigenvalue problem arising in models of enhanced oil recovery}
\title{On an eigenvalue problem arising in a Hele-Shaw model of enhanced oil recovery: The case of linear viscous profile}
\title{On a three-layer Hele-Shaw model of enhanced oil recovery with a linear viscous profile}

\author{
Prabir Daripa\footnote {Author for correspondence (e-mail:
prabir.daripa@math.tamu.edu)}\\
Department of Mathematics\\Texas A\&M University\\
College Station, Texas\vspace{0.5cm}\\\\
Oscar Orellana and Rodrigo Meneses\\
Departamento de Matem\'aticas\\
Universidad T\'ecnica Santa Mar\'ia de Valpara\'iso\\
UTFSM, Chile}
\maketitle

\begin{abstract}
We present a non-standard eigenvalue problem that arises in the linear stability of a three-layer Hele-Shaw model of enhanced oil recovery. A nonlinear transformation is introduced which allows reformulation of the non-standard eigenvalue problem as a boundary value problem for Kummer's equation when the viscous profile of the middle layer is linear. Using the existing body of works on Kummer's equation, we construct an exact solution of the eigenvalue problem and provide the dispersion relation implicitly through the existence criterion for the non-trivial solution. We also discuss the convergence of the series solution. It is shown that this solution reduces to the physically relevant solutions in two asymptotic limits: (i) when the linear viscous profile approaches a constant viscous profile; or (ii) when the length of the middle layer approaches zero.
\end{abstract}

\smallskip

\newpage

\section{Introduction}\label{introduction}

The flow of two immiscible fluids through porous media arises in many important industrial and natural situations such as secondary oil recovery, ground water remediation, and geological ${\rm CO}_2$ storage. Such flows are known to be potentially unstable, especially when the displacing fluid is more viscous than the displaced one. There exist some similarities between porous media and Hele-Shaw flows (i.e. flow in a Hele-Shaw cell, see below); for example the pressure drop in both such flows are governed by Darcy's law for single fluid flow. Due to this and the fact that it is significantly easier to study Hele-Shaw flows theoretically, numerically, and experimentally, there have been numerous theoretical and numerical studies even for Hele-Shaw flow of two immiscible fluids since the early 1950s, starting with the work of Saffman and Taylor~\cite{Saffman/Taylor:1958}. There are many review articles on such studies, for example see \cite{Homsy:1987, Saffman:1986}. These studies were originally motivated by displacement processes arising in secondary oil recovery, even though these studies have much wider appeal in the sciences and engineering. In the late 1970s, tertiary displacement processes involved in chemical enhanced oil recovery generated interest in three-layer and multi-layer Hele-Shaw flows (see \cite{daripa08:multi-layer, daripa08:studies, dp05:growth, GH83:optimal}).

In this paper, we first briefly derive the non-standard eigenvalue problem. This eigenvalue problem has been derived earlier by the first author and his collaborators; for example see~\cite{daripa08:multi-layer}. But the difference is that the derivation presented here is more general and shows how to generate higher order correction terms if necessary in order to study the effect of nonlinear terms that may dominate the dynamics, particularly in view of the sensitivity of fingering problems to finite amplitude perturbations. However, we do not study or discuss such nonlinear effects in this paper which will be taken up in the future as it falls outside the scope of this paper. We then analytically study this non-standard eigenvalue problem using non-linear transformation for the case when the viscous profile of the middle layer is linear. We will see below that this case is relatively hard to study in comparison to the case when the viscous profile is exponential which we have recently addressed in \cite{gin-daripa:hs-rect}.

\begin{figure}[ht]
\centering
\includegraphics[scale=.8]{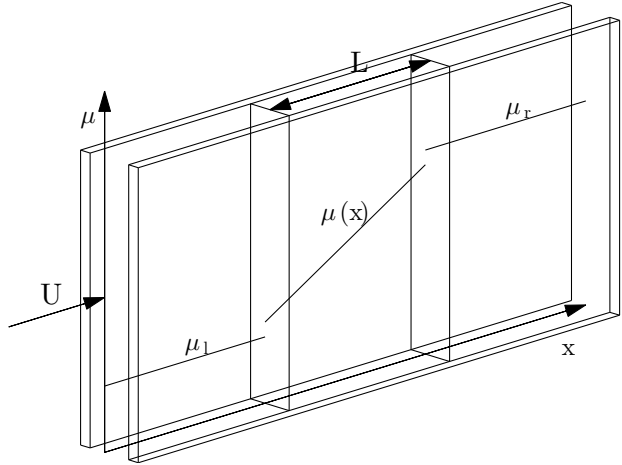}
\caption{Three-layer rectilinear Hele-Shaw flow in which the middle layer has a smooth viscous profile. The physical set-up as well as the smooth viscous profile of the middle layer are shown in this figure.}
\label{fig:HS_Rect_LinearProfile}
\end{figure}

The physical set-up consists of rectilinear motion of three immiscible fluids in a Hele-Shaw cell which is a device separating two parallel plates by a distance $b$ (see Fig.~\ref{fig:HS_Rect_LinearProfile}). The fluid in the extreme left layer $R_1$ with viscosity $\mu_1$ extends up to $x=-\infty$, the fluid in the extreme right layer $R_2$ with viscosity $\mu_2 > \mu_1$ extends up to $x=\infty$, and the fluid in the middle-layer $R_I$ of finite length $L$ has a smooth viscous profile with viscosity increasing in the direction of displacement. The interfacial tensions of the leading and the trailing interfaces are given by $T$ and $S$ respectively. It is well established that this Hele-Shaw flow is similar to flow in homogeneous porous media with equivalent permeability $b^2/12$. Without any loss of generality, we take this to be one below. The mathematical model considered here consists of conservation of mass, Darcy's law and advection equation for viscosity. Thus we have
\begin{align}\label{eq1}
&{\bf \nabla}\cdot{\bf u}=0,\quad \forall\ (x,y)\in {\bb R}^2\\
\label{eq2}
&{\bf \nabla}{p}=-\mu{\bf u}\ \quad \forall\ (x,y)\in {\bb R}^2, \text{ but } \mu(x,y,t) = \left\{\begin{array}{l} \mu_1; \text{ if } (x,y)\in R_1\\ \mu(x,y,t); \text{ if } (x,y)\in R_I\\ \mu_2; \text{ if } (x,y)\in R_2\end{array}\right.\\
\label{eq3}
&\mu_t + u \mu_x + v\mu_y = 0;\quad \forall\ x\in R_I\ \text{and}\ y\in {\bb R}.
\end{align}
Due to the continuity equation, we can define the stream function $\psi = \psi(x,y,t)$ such that $u=\psi_y$ and $v=-\psi_x$. This then implies that
\begin{equation}
p_x = -\mu\psi_y,\quad p_y = \mu\psi_x,\quad {\rm and}\quad \mu_t + \psi_y \mu_x - \psi_x\mu_y = 0.
\end{equation}
Since ${\bf u} = (U_0,0)$ when $x^2+y^2\to \infty$, we consider a small perturbation of the basic scalar fields $\psi_0,p_0$ and $\mu_0$ of the form
\begin{equation*}
\left.
\begin{array}{l l}
\psi &= \psi(x,y,t) = U_0y + \vp\widehat\psi(x,y,t)\\
p &= p(x,y,t) = p_0(x,t) + \vp\widehat p(x,y,t)\\
\mu &=\mu(x,y,t) = \mu_0(x,t) + \vp\widehat\mu(x,y,t)
\end{array}
\right\}
\end{equation*}
Substituting into the original equations, we get the following $O(\vp^0)$ and $O(\vp^1)$ equations.\newline
\n $O(\vp^0)$ equations:
\begin{equation*}
\left.
\begin{array}{l l}
 &p_{0x} = -U_0\mu_0\\
&p_{0y} = 0\\
&\mu_{0t} + U_0\mu_{0x}= 0
\end{array}
\right\}
\end{equation*}
These equations provide the basic solution given by
\begin{equation*}
\left.
\begin{array}{l}
\mu_0 = \mu_0(x,t) = \mu_0(x-U_0t)\\
p_0 = p_0(x,t) = -U_0 {\ds\int^x_{x_0}} \mu_0(s-U_0t)ds\\
{\bf u} = (U_0,0)
\end{array}
\right\}
\end{equation*}
where $\mu_0(x-U_0t)$ is an arbitrary function of $\xi = x-U_0t$, meaning the viscous profile is fixed with respect to a moving frame moving at a constant velocity $(U_0,0)$.\ms
\newline
\n $O(\vp^1)$ equations:
\begin{equation*}
\left.
\begin{array}{l l}
&\widehat p_x = -U_0\widehat\mu - \mu_0\widehat\psi_y\\
&\widehat p_y = \mu_0\widehat\psi_x\\
&\widehat\mu_t + U_0\widehat\mu_x + \mu_{0x}\widehat\psi_y = 0.
\end{array}
\right\}
\end{equation*}
Now, introducing the moving frame change of variables, namely $\xi = x - U_0t,\ y = y,\ t = t$, we get the following system of equations.
\begin{equation}\label{eqn:Eqq1}
\left.
\begin{array}{l l}
 &\widehat p_\xi = - U_0\widehat\mu - \mu_0\widehat\psi_y\\
&\widehat p_y = \mu_0\widehat\psi_\xi\\
&\widehat\mu_t + \mu_{0\xi}\widehat\psi_y = 0.
\end{array}
\right\}
\end{equation}
Taking cross derivatives of the first two equations with respect to $y$ and $\xi$ respectively and then subtracting the resulting equations from each other gives
$\mu_{0\xi}\widehat\psi_\xi + U_0\widehat\mu_y + \mu_0\Delta\widehat\psi=0.$ This combined with the equation $\eqref{eqn:Eqq1}_3$ leads to
\begin{equation*}
\mu_{0\xi}\widehat\psi_{\xi t} - U_0\mu_{0\xi} \widehat\psi_{yy} + \mu_0(\Delta\widehat\psi)_t = 0.
\end{equation*}
Using the ansatz $\widehat\psi=f(\xi) e^{iky+\sigma t}$ in the above equation together with the appropriate boundary conditions (see \cite{daripa08:multi-layer}) give the following eigenvalue problem.
\begin{align*}
 &\mu_0(\xi)\{f_{\xi\xi} - k^2f\} + \mu_{0\xi}(\xi) f_\xi + \frac{k^2U_0}\sigma \mu_{0\xi}(\xi) f=0\\
&\mu^+_0(-L)f_\xi(-L) = f(-L)\left\{\mu_1k + \frac{U_0k^2}\sigma [\mu_1-\mu^+_0(-L)] + \frac{Sk^4}\sigma\right\}\\
&\mu^-_0(0)f_\xi(0) = f(0) \left\{-\mu_2k + \frac{U_0k^2}\sigma [\mu_2-\mu^-_0(0)] - \frac{Tk^4}\sigma\right\}
\end{align*}
where the viscous profile of the middle layer, namely, $\mu_0=\mu_0(\xi)$ is an arbitrary function. This is a non-standard eigenvalue problem in that the spectral parameter $1/\sigma$ appe
ars in the equation as well as in the boundary conditions. Recently, this problem has been numerically solved by Daripa~\cite{daripa08:studies} for a constant viscous profile and by Daripa \& Ding~\cite{daripa:tipm2012} for non-constant viscous profiles $\mu_0(\xi)$ to determine the most optimal profile, i.e., the least unstable profile. This problem has been too difficult to solve analytically for non-constant profiles. Progress made in this direction for the linear viscous profile is presented below.

In this paper we consider a linear viscous profile for the intermediate fluid region given by
$$ \mu_{0}(\xi)=\alpha \xi +\beta\qquad \textrm{for}\ -L<\xi<0  $$
 where
$$ \alpha=\frac{ (\mu_{2}-\mu_{1})-(J_{1}+J_{2}) }{L}=\frac{\mu_0(0)-\mu_0(-L)}{L},\qquad \beta=\mu_{2}-J_{2}=\mu_0(0),   $$
and $J_{1}=\mu_{0}(-L)-\mu_{1}, J_{2}=\mu_{2}-\mu_{0}(0)$ are jump discontinuity values at the interfaces $\xi=-L$ and $\xi=0$, respectively. In the left region the problem reduces to
$f_{\xi\xi} - k^2f = 0, \lim_{\xi\to-\infty} f(\xi) = 0,$ which has solution $f(\xi) = f(-L)e^{k(\xi+L)} \text{ for } \xi<-L$.
In the right region the problem reduces to
$f_{\xi\xi} -k^2f = 0, \lim_{\xi\to\infty} f(\xi)=0,$ which has solution $f(\xi) = f(0)e^{-k\xi} \text{ for } \xi>0.$
In the intermediate region the problem reduces to
\begin{equation}\label{eq:main-evp}
  \left\lbrace
              \begin{array}{rcl}
              \displaystyle{(\alpha\xi +\beta)f_{\xi\xi}+\alpha f_{\xi}+k^{2}(\alpha \lambda -(\alpha\xi+\beta))f}&=&0,\quad -L<\xi<0\\
              \displaystyle{      \mu_{0}(-L)f_{\xi}(-L)}                                 &=& (\alpha_{1}(k)\lambda+\alpha_{2}(k))f(-L) \\
              \displaystyle{  \mu_{0}(0)f_{\xi}(0)  }                                    &=& (\beta_{1}(k)\lambda-\beta_{2}(k))f(0)
              \end{array}
  \right.
 \end{equation}
where $\lambda=\frac{U_{0}}{\sigma}$ is the spectral parameter and
\begin{equation}\label{coef}
  \left\lbrace
              \begin{array}{cc}
              \alpha_{1}(k)=\frac{Sk^{4}}{U_{0}} +k^{2}(\mu_{1}-\mu_{0}(-L)),  & \alpha_{2}(k)=\mu_{1}k\\
              \beta_{1}(k) = -\frac{T k^{4}}{U_{0}}+ k^2(\mu_{2}-\mu_{0}(0)),  & \beta_{2}(k)=\mu_{2}k
              \end{array}
  \right.
  \end{equation}

\section{Solution to the Eigenvalue Problem~\eqref{eq:main-evp} via Kummer's Equation}
We introduce the nonlinear transformation and change of variables given by
\begin{equation}\label{eq:change-of-variables}
 f(\xi) = e^{k\xi}z(w),\quad w=-2k\left(\xi+\frac{\beta}{\alpha}\right) < 0. 
\end{equation}
After some manipulation of the eigenvalue problem \eqref{eq:main-evp} using the above transformation, we obtain the following eigenvalue problem for the Kummer's equation~$\eqref{eq:bc-z-Kummer}_1$.
\begin{equation}\label{eq:bc-z-Kummer}
\left.
\begin{array}{ll}
&wz_{ww} + (b-w)z_w - az = 0,\quad w_{2}<w<w_{1}\\
&\eta_1z(w_1) + \phi_1z^{\prime}(w_1) = 0,\\
&\eta_2z(w_2) + \phi_2z^{\prime}(w_2) = 0,
\end{array}
\right\}
\end{equation}
where a prime denotes derivative,
\begin{equation}\label{eq:w1-w2-def}
\left.
\begin{split}
b=1,\quad a=\frac{1}{2}(1+k\lambda),\quad w_{2}\equiv w(\xi=0)=-2k\ \frac{\mu_{0}(0)}{\alpha},\quad w_{1} \equiv w(\xi=-L)=w_{2}+2kL<0,\\
\eta_1 = {Sk^4}/\sigma,\quad \phi_1=2k\mu_1,\quad \eta_2 = 2\mu_2k+{Tk^4}/\sigma,\quad {\rm and}\quad \phi_2=-2k\mu_2.
\end{split}
\right\}
\end{equation}
The eigenvalue problem \eqref{eq:bc-z-Kummer} is a regular two point boundary value problem for each wave number $k$. One solution $z_1$ of the Kummer's equation is given by 
\begin{equation*}
z_1(a,1,w) = a_0+\frac{a_1w}{(1!)^2} + \frac{a_2w^2}{(2!)^2} + \frac{a_3w^3}{(3!)^2} +\cdot + \frac{a_nw^n}{(n!)^2} +\cdots.
\end{equation*}
\newline where $a_0=1$ and $a_n = a(a+1)(a+2)\cdots (a+n-1)\ \ {\rm for}\ n=1,2,\ldots$.\newline This is an analytic solution. It is easily seen that the derivative of this solution which we will need below for the dispersion relation is given by
{\small
\begin{equation*}
 z'_1(a,1,w) = \frac{a_1}{(1!)^2} + \frac{2a_2w}{(2!)^2} + \frac{3a_3w^2}{(3!)^2} +\cdots+ \frac{na_nw^{n-1}}{(n!)^2} + \frac{(n+1)a_{n+1}w^n}{(n+1)!^2} +\cdots = az_1(a+1,2,w).
\end{equation*}
}
The linearly independent second solution is easily constructed by the method of Frobenius. Avoiding all the details, the second solution $z_2$ is given by
{\small
\begin{equation*}
z_2(a,1,w) = w^{1/2} \left\{1 + \frac{a_{1/2}w}{1\cdot 3\cdot 1!} + \frac{a_{3/2}w^2}{1\cdot 3\cdot 5\cdot 2!} + \frac{a_{5/2}w^3}{1\cdot 3\cdot 5\cdot 7\cdot 3!} +\cdots+\frac{a_{\frac{2n-1}2}w^n}{1 \cdot 3\cdot 5\cdot 7\cdots(2n+1)\cdot n!} +\cdots\right\}
\end{equation*}
}
where $a_{{}_{\frac{2n-1}2}}=(a+\frac{2n-1}2)(a+\frac{2n-3}2) \cdots (a+\frac52)(a+\frac32)(a+\frac12),\quad n=1,2,3,\ldots$~. Its derivative which we will need below is then given by
{\small
\begin{equation*}
z'_2(a,1,w) = \frac{1}{2w}z_2(a,1,w) + w^{1/2}
\left\{\frac{2a_{1/2}}{3!} + \frac{2\cdot 2^2a_{3/2}w}{5!}+ \frac{3\cdot 2^3 a_{\frac52}w^2}{7!} +\cdots+
\frac{n2^n a_{{}_{\frac{2n-1}2}} w^{n-1}}{(2n+1)!}+\cdots\right\}
\end{equation*}
}
The general solution of the Kummer's equation is then given by
\begin{equation*}
z(w) = c_1z_1(a,1,w) + c_2z_2(a,1,w),
\end{equation*}
where $c_1$ and $c_2$ are arbitrary constants. Substituting the general solution into the two boundary conditions of the eigenvalue problem \eqref{eq:bc-z-Kummer}, we obtain the following linear system of equations for $c_1$ and $c_2$.
\begin{align*}
&\{\eta_1z_1(w_1) + \phi_1z'_1(w_1)\} c_1 + \{\eta_1z_2(w_1) + \phi_1z'_2(w_1)\} c_2 = 0,\\
&\{\eta_2z_1(w_2) + \phi_2z'_1(w_2)\} c_1 + \{\eta_2z_2(w_2) + \phi_2z'_2(w_2)\} c_2 = 0.
\end{align*}
Therefore, for a non-trivial solution we have
\begin{equation}\label{eqn:existence}
\begin{vmatrix}
\eta_1z_1(w_1) + \phi_1z'_1(w_1)&\eta_1z_2(w_1) + \phi_1z'_2(w_1)\\
\eta_2z_1(w_2) + \phi_2z'_1(w_2)&\eta_2z_2(w_2) + \phi_2z'_2(w_2).
\end{vmatrix}
=0.
\end{equation}
This formally gives the dispersion relation $\sigma(k)$ in terms of the problem data: $S, T, \mu_1, \mu_2, L$ and $U_0$. 

In terms of the original variables $f$ and $\xi$, the fundamental solutions $f_1(\xi)$ and $f_2(\xi)$  are then given by (see \eqref{eq:change-of-variables})
\begin{align}
 f_1(\xi) &= e^{k\xi}\left\{1-\frac{a_12k(\xi+\frac{\mu_2}\alpha)}{(1!)^2} +
\frac{a_2(2k)^2(\xi + \frac{\mu_2}\alpha)^2}{(2!)^2}-
\frac{a_3(2k)^3(\xi +\frac{\mu_2}\alpha)^3}{(3!)^2} +\right.\nonumber\\
&\quad \left. \cdots+ (-1)^n \frac{a_n(2k)^n(\xi +\frac{\mu_2}\alpha)^n}{(n!)^2} +\cdots\right\}\label{eqn:f1}\\
f_2(\xi) &= e^{k\xi}\left\{-2k\left(\xi + \frac{\mu_2}\alpha\right)\right\}^{1/2}
\left\{1-\frac{a_{1/2}2k(\xi + \frac{\mu_2}\alpha)}{1\cdot 3\cdot(1!)} +
\frac{a_{3/2}(2k)^2 (\xi+\frac{\mu_2}\alpha)^2}{1\cdot 3\cdot 5\cdot (2!)}\right.\nonumber\\
&\quad \left. - \frac{a_{5/2}(2k)^3(\xi+\frac{\mu_2}\alpha)^3}{1\cdot 3\cdot 5\cdot 7\cdot (3!)}
+\cdots+ (-1)^n \frac{a_{{}_{\frac{2n-1}2}}(2k)^n (\xi+\frac{\mu_2}\alpha)^2}
{1\cdot 3\cdot 5\cdot 7\cdots(2n+1)(n!)} +\cdots\right\}\label{eqn:f2}
\end{align}
where
\[
\left.
\begin{array}{l}
a_0=1\\
a_n = a(a+1)(a+2)\cdots (a+n-1)\\
a_{\frac{2n-1}2} = (a+1/2)(a+3/2)(a+5/2)\cdots \left(a+\frac{2n-3}2\right) \left(a + \frac{2n-1}2\right),\quad 1,2,3,4,5,\ldots\\
a_{-\frac12} = 1
\end{array}
\right\}
\]
Also, recall that $\xi = x-U_0t$, $\alpha = (\mu_2-\mu_1)/L$ and $a = (1+kU_0/\sigma)/2$. Noticing that both series are centered at $\xi = -(\mu_2/\alpha)$ and applying the ratio test for series for $f_1(\xi)$ above, we have
\begin{equation*}
\left|\frac{\dfrac{a_{n+1}\cancel{(2k)^{n+1}} \cancel{(\xi+\frac{\mu_2}\alpha)^{n+1}}}{((n+1)!)^2}}{\dfrac{a_n \cancel{(2k)^n} \cancel{(\xi+\frac{\mu_2}\alpha)^n}}{(n!)^2}}\right| = \left|\frac{(a+n)(2k)(\xi +\frac{\mu_2}\alpha)}{(n+1)^2}\right|
= \left|\frac{(a+n)(2k)}{(n+1)^2}\right| \left|\xi+\frac{\mu_2}\alpha\right| ~
\rightarrow ~ 0\ \ \text{as}\ n\rightarrow \infty,
\end{equation*}
for any $\xi$ fixed. Thus the series for $f_1(\xi)$ converges absolutely $\forall\ \xi \ne -(\mu_2/\alpha)$, but the series \emph{evaluated} at $\xi = -(\mu_2/\alpha)$ reduces to 1. This implies that the radius of this series is $\infty$. Hence
\[
 \lim_{n\to\infty} \frac{(-1)^n a_n(2k)^n(\xi+\frac{\mu_2}\alpha)^n}{(n!)^2} = 0,
\]
and since it is an alternating singular series, the error $E_n^{(1)}$ in approximating $f_1$ by terms up to $k^n$ is smaller than the last neglected term, namely
\[
 |E_n^{(1)}| \leqq \left|\frac{a_{n+1}(2k)^{n+1}(\xi+\frac{\mu_2}\alpha)^{n+1}}{((n+1)!)^2}\right|
\]
for a fixed $\xi$. Similarly, applying the ratio test to the series for $f_2(\xi)$,
\begin{equation*}
\left|\frac{\dfrac{\cancel{2^{n+1}} \cancel{a_{{}_{\frac{2n+1}2}}} \cancel{(2k)^{n+1}} (\xi + \frac{\mu_2}\alpha)^{n+1}}{(2n+3)!}}{\dfrac{\cancel{2^n} \cancel{a_{{}_{\frac{2n-1}2}}} \cancel{(2k)^n} (\xi + \frac{\mu_2}\alpha)^n}{(2n+1)!}}\right| = \left|\frac{2(a+ \frac{2n+1}2) (2k) (\xi + \frac{\mu_2}\alpha)}{(2n+2)(2n+3)}\right| \to 0,\ \text{as}\ n \rightarrow \infty,
\end{equation*}
for any $\xi$ \text{fixed}. Thus the series inside the brackets converges absolutely $\forall \xi\ne -(\mu_2/\alpha)$, but the series inside the brackets evaluated at $\xi = -\frac{\mu_2}\
alpha$ reduces to 1. Therefore, the radius of convergence of the series within the brackets is $\infty$. Hence
\begin{equation*}
\lim\limits_{n\to\infty}\frac{(-1)^n 2^n a_{{}_{\frac{2n-1}2}}(2k)^n (\xi+\frac{\mu_2}\alpha)^n}{(2n+1)!} = 0.
\end{equation*}
Notice that $f_2(\xi)$ has a branch point at $\xi=-\frac{\mu_2}\alpha$. In any case, since the series within the brackets is an alternating sign series, if we truncate it, the error $E_n^{
(2)}$ is smaller than the last neglected term, i.e.,
\[
 |E_n^{(2)}| \leqq \left|\frac{2^{n+1}a_{{}_{\frac{2n+1}2}}(2k)^{n+1} (\xi+ \frac{\mu_2}\alpha)^{n+1}}{(2n+3)!}\right|.
\]
The general solution of the ODE $\eqref{eq:main-evp}_1$ is then given by $f(\xi)=c_1f_1(\xi)+c_2f_2(\xi)$. The boundary values of $f(\xi)$ follow from \eqref{eqn:f1} and \eqref{eqn:f2} whi
ch are now given by
\begin{align*}
 f_1(-L) &= e^{-kL} \sum^\infty_{n=0} \frac{(-1)^n (2kL)^n a_n (\frac{\mu_1}{\mu_2-\mu_1})^n}{(n!)^2}\\
f_1(0) &= \sum^\infty_{n=0} \frac{(-1)^n (2kL)^n a_n (\frac{\mu_2}{\mu_2-\mu_1})^n}{(n!)^2}\\
f_2(-L) &= e^{-kL} \left\{-2kL \left(\frac{\mu_1}{\mu_2-\mu_1}\right)\right\}^{1/2} \sum^\infty_{n=0} \frac{(-1)^n (2^2kL)^n a_{{}_{\frac{2n-1}2}}(\frac{\mu_1}{\mu_2-\mu_1})^n}{(2n +1)!}\\
f_2(0) &= \left\{-2kL \left(\frac{\mu_2}{\mu_2-\mu_1}\right)\right\}^{1/2} \sum^\infty_{n=0} \frac{(-1)^n (2^2kL)^n a_{{}_{\frac{2n-1}2}}(\frac{\mu_2}{\mu_2-\mu_1})^n}{(2n+1)!}
\end{align*}
Substituting these in the boundary conditions $\eqref{eq:main-evp}_3$, we obtain the following system of equations for the constants $c_1$ and $c_2$.
\begin{align*}
\{A_1f_1(-L) - \mu_1f'_1(-L)\}c_1 + \{A_1f_2(-L) - \mu_1f'_2(-L)\} c_2 &= 0,\\
\{A_2f_1(0) - \mu_2f'_1(0)\}c_1 + \{A_2f_2(0) - \mu_2f'_2(0)\} c_2 &= 0,
\end{align*}
where $A_1 = \{\mu_1k + \frac{Sk^4}\sigma\}$ and $A_2 = \{-\mu_2k+ \frac{Tk^4}\sigma\}$. For the existence of nontrivial solutions, we then have
\[
 \begin{vmatrix}
  A_1f_1(-L)-\mu_1f'_1(-L)&A_1f_2(-L)-\mu_1f'_2(-L)\\
A_2f_1(0)-\mu_2f'_1(0)&A_2f_2(0)-\mu_2f'_2(0)
 \end{vmatrix} = 0,
\]
which gives us the dispersion relation in the form: $\theta(\sigma,k)=0$. Because of the nature of the series solutions given above, it is not possible to give this dispersion relation explicitly. 

\section{Limiting Cases}\label{section:limiting-cases}

There are an infinite number of eigenvalues $\sigma$ (recall $\lambda=U/\sigma$) which can be ordered: $\sigma_{\max}=\sigma_1 > \sigma_2 > .....> \sigma_\infty \rightarrow 0$. We know that these infinite number of eigenvalues should reduce to (i) only two in the limit $\alpha \to 0$ corresponding to the constant viscosity of the intermediate layer fluid~(see Daripa~\cite{daripa08:studies}); (ii) only one in the limit $L \to 0$ (see Saffman \& Taylor~\cite{Saffman/Taylor:1958}, Daripa~\cite{daripa08:multi-layer})  and (iii) only two in the limit of $L \to \infty$ (see Daripa~\cite{daripa:tipm2012}). In fact, we also know the eigenvalues in these limiting cases from the pure Saffman-Taylor growth rate of individual interfaces. These results by no means are transparent from the solutions of the eigenvalue problem~\eqref{eq:main-evp} given in the previous section. Below, we show how to recover these limit solutions (eigenvalues) from the infinite number of eigenvalues for the linear viscous profile.

\subsection{Constant viscosity case: $\alpha=0$.}\label{subsection:constant-viscosity}
In this case, the eigenvalue problem~\eqref{eq:main-evp} reduces to
\begin{equation}\label{SLP3}
\left\lbrace
\begin{array}{rcl}
f_{\xi\xi}-k^{2}f                     &=&0,\quad -L<\xi<0,\\
\mu_{0}(-L)f_{\xi}(-L) &=& (\alpha_{1}(k)\lambda+\alpha_{2}(k))f(-L), \\
\mu_{0}(0)f_{\xi}(0)   &=& (\beta_{1}(k)\lambda-\beta_{2}(k))f(0),
\end{array}
\right.
\end{equation}
In this case, the change of variable introduced previously, namely $w=-2k\left(\xi+\frac{\beta}{\alpha}\right)$, which converts the equation~$\eqref{eq:main-evp}_1$ to Kummer's equation, i
s not well-defined. Therefore we work with the boundary value problem~$\eqref{SLP3}$. Now, consider the general solution of the ODE in (\ref{SLP3})
$$f(\xi)=Af_{1}(\xi)+Bf_{2}(\xi),$$
such that
\begin{equation}\label{cond1}
\left\lbrace\begin{array}{cc}
f_{1}(-L)=1, & f_{1}'(-L)=0,\\
f_{2}(-L)=0, & f_{2}'(-L)=1.
\end{array}
\right.
\end{equation}
Therefore, we get
\begin{equation*}
f_{1}(\xi)=\cosh(k(\xi+L)),\quad\textrm{and}\quad f_{2}(\xi)=\frac{\sinh(k(\xi+L))}{k}.
\end{equation*}
We search for a solution of the boundary value problem (\ref{SLP3}) of the form
\begin{equation}\label{eq:limf}
f(\xi;\lambda)=A\cosh(k(\xi+L)) +  B \frac{\sinh(k(\xi+L))}{k}.
\end{equation}
To find a solution of (\ref{SLP3}) of this form, we start determining the coefficients using the shooting technique such that the boundary condition at $\xi=-L$ is satisfied. Obviously, the coefficients $A$ and $B$ depend on the parameter $\lambda$. Then, we find $\lambda$ in such a way that the solution satisfies the boundary condition at $\xi=0$. Hence, we look for $A$ and $B$ such that
\begin{equation*}
f(-L;\lambda)=\mu_{0}(-L),\quad \textrm{and}\quad f_{\xi}(-L;\lambda)=\alpha_{1}(k)\lambda + \alpha_{2}(k).
\end{equation*}
Then it follows directly from (\ref{cond1}) that $A= \mu_{0}(-L)$ and $B=\alpha_{1}(k)\lambda + \alpha_{2}(k).$ Therefore
\begin{equation}\label{solmodelosimple}
f(\xi;\lambda)=\mu_{0}(-L)\cosh(k(\xi+L))+(\alpha_{1}(k)\lambda + \alpha_{2}(k)) \frac{\sinh(k(\xi+L))}{k}
\end{equation}
satisfies the ODE in \eqref{SLP3} and the boundary condition at $\xi=-L$. From these it follows that the spectrum of problem (\ref{SLP3}) can be studied using the following algebraic equation (see $\eqref{SLP3}_3$)
\begin{equation}\label{al1}
\mu_{0}(0)\frac{f_{\xi}(0;\lambda)}{f(0;\lambda)}=\beta_{1}(k)\lambda -\beta_{2}(k),
\end{equation}
where $f(\xi;\lambda)$ is the function defined in \eqref{solmodelosimple}. Evaluating $f(0;\lambda)$ and $f_{\xi}(0;\lambda)$ from \eqref{solmodelosimple} and substituting directly in \eqref{al1} one obtains
\begin{equation}\label{explicitall}
k\mu_{0}(0)\left\lbrace \frac{\mu_{0}(-L)\sinh(kL) +  (\alpha_{1}(k)\lambda + \alpha_{2}(k)) \displaystyle{\frac{\cosh(k L )}{k}}}{\mu_{0}(-L)\cosh(kL) +  (\alpha_{1}(k)\lambda + \alpha_{2}(k)) \displaystyle{\frac{\sinh(kL)}{k}}}\right\rbrace=\beta_{1}(k)\lambda -\beta_{2}(k).
\end{equation}
Then, taking $L\to 0^{+}$ one obtains
\begin{equation*}
k\mu_{0}(0)\left( \frac{\alpha_{1}(k)\lambda+\alpha_{2}(k)}{k\mu_{0}(-L)}  \right) =  \beta_{1}(k)\lambda -\beta_{2}(k),
\end{equation*}
which is equivalent to the equation
\begin{equation*}
\left(\mu_{0}(0)\alpha_{1}(k)- \mu_{0}(-L)\beta_{1}(k)\right)\,\lambda = -\left(\mu_{0}(0)\alpha_{2}(k)+ \mu_{0}(-L)\beta_{2}(k)\right).
\end{equation*}
Since $\mu_{0}(\xi)={\rm constant}$ ($\mu$ or $\beta$), it follows that
\begin{equation*}
(\alpha_{1}(k)- \beta_{1}(k)) \lambda =- (\alpha_{2}(k)+ \beta_{2}(k)).
\end{equation*}
Now, using the definition of the coefficient $\alpha_{1}(k),\ \alpha_{2}(k),\ \beta_{1}(k),\ \beta_{2}(k)$ given in (\ref{coef}) we have
\begin{equation*}
\left[\left(\frac{Sk^{4}}{U_{0}} +\frac{Tk^{4}}{U_{0}}\right)+k^{2}(\mu_{1}-\mu_{2})\right] \lambda=-k(\mu_{1}+\mu_{2})
\end{equation*}
from which it follows that
\begin{equation*}
\sigma=\frac{U_{0}k(\mu_{2}-\mu_{1})-k^3(S+T)}{(\mu_{1}+\mu_{2} )}
\end{equation*}
which is the formula for the growth rate of an interface with surface tension $(S+T)$, which is what should be expected in this limit. Thus we recover the classical formula for the growth rate in this limit.

To take the limit when $L\to\infty$, we go back to equation (\ref{explicitall}) and write it as follows
\begin{equation*}
k\mu_{0}(0)  \left\lbrace   \frac{\mu_{0}(-L)\tanh(k L) +  (\alpha_{1}(k)\lambda + \alpha_{2}(k)) \displaystyle{\frac{1}{k}}}{\mu_{0}(-L) +  (\alpha_{1}(k)\lambda + \alpha_{2}(k)) \displaystyle{\frac{\tanh(k L)}{k}}}    \right\rbrace=\beta_{1}(k)\lambda -\beta_{2}(k).
\end{equation*}
Now, taking the limit when $L\to \infty$, we obtain $k\mu_{0}(0) = \beta_{1}(k)\lambda -\beta_{2}(k).$ Using $\sigma=U_{0}/\lambda$ and expressions for the coefficients from $\eqref{coef}_2$, we obtain
\begin{equation*}
\sigma=U_{0}\left( \frac{-\frac{Tk^{4}}{U_{0}} +k^{2}(\mu_{2}-\mu_{0}(0))}{k(\mu_{2}-\mu_{0}(0))}\right).
\end{equation*}
Finally, since $\mu_{0}(0)={\rm constant}$ ($\mu$ or $\beta$) it follows that
\begin{equation*}
\sigma=-\frac{Tk^{3}}{\mu_{2}+\mu}  +U_{0}k \left( \frac{\mu_{2}-\mu}{\mu+\mu_{2}}  \right)
\end{equation*}
which gives the classical formula for Saffman-Taylor instability of the leading interface. Similarly, we can recover the the classical formula for Saffman-Taylor instability of the trailin
g interface by reversing the shooting technique (see after \eqref{eq:limf}), i.e., first find the solution which is analogous to \eqref{solmodelosimple} but satisfies the boundary condition 
at $\xi=0$ instead and then shoot to satisfy the boundary condition at $\xi=-L$ (i.e., replace \eqref{al1} by a similar formula derived from the boundary condition at $\xi=-L$ and follow the
 procedure).

\subsection{Linear viscosity case: $\alpha > 0$.}\label{subsection:linear-viscosity}

In this section, we study asymptotic limits ($L\to0$ and $L\to\infty$) of the solutions to the eigenvalue problem~\eqref{eq:main-evp}. To this end, we consider the following form of two linearly independent solutions of Kummer's equation $(\ref{eq:bc-z-Kummer})_1$.
These are convenient for the asymptotic analysis presented below.
\begin{equation}\label{specialfunctions}
\left.
\begin{array}{rcl}
M(a,1,w)         &=&\displaystyle{1+\sum_{i=1}^{\infty}\frac{(a)_{i}}{(1)!}\frac{w^{i}}{i!}}\\
e^{w}U(1-a,1,-w) &=&\displaystyle{ -\frac{e^{w}}{\Gamma(1-a)}  M(1-a,1,-w)\ln(-w)+ } \\
&  &  \displaystyle{ -\frac{e^{w}}{\Gamma(1-a)} \sum_{i=1}^{\infty} \frac{(1-a)_{i}}{(i!)^{2}}\left( \Psi((1-a)+i)-2\Psi(1+i)    \right)(-w)^{i}         }
\end{array}
\right\}
\end{equation}
where $\Psi(s)$ is Euler's digamma function (See Abramowitz~\cite{Abramowitz:1964}, Chapter 13).

To this end, we follow the steps presented in the previous section~\ref{subsection:constant-viscosity} for the particular case $\mu_{0}(\xi)={\rm constant}$ ($\mu$ or $\beta$). From the transformation in \eqref{eq:change-of-variables}, it follows that
\begin{equation}\label{eq:fxi}
f(\xi;\lambda)=(Az_{1}(w)+Bz_{2}(w))e^{k\xi}
\end{equation}
is the general solution of the ODE $\eqref{eq:main-evp}_1$ where
\begin{equation}\label{eq:two-lin}
\left.
\begin{array}{rcl}
z_{1}(w)&=& C_{1}M(a,1,w)+D_{1}e^{w}U(1-a,1,-w)\\
z_{2}(w)&=& C_{2}M(a,1,w)+D_{2}e^{w}U(1-a,1,-w)
\end{array}
\right\}
\end{equation}
and $C_{1},\ D_{1},\ C_{2},\ D_{2} $  are chosen such that
\begin{equation}\label{eq:cond2}
\left.
\begin{array}{cc}
z_{1}(w_{1})=1, & z_{1}'(w_{1})=0\\
z_{2}(w_{1})=0, & z_{2}'(w_{1})=1
\end{array}
\right\}
\end{equation}
where
$a=(1+k\lambda)/2$.
Substituting \eqref{eq:two-lin} in the boundary conditions~\eqref{eq:cond2}, we obtain the following linear systems of equations
\begin{equation}\label{eq:system1}
\begin{pmatrix}
M(a,1,w_{1})&e^{w_{1}} U(1-a,1,-w_{1})\\M'(a,1,w_{1})& (e^{w} U(1-a,1,-w)'_{w_{1}}
\end{pmatrix}
\begin{pmatrix}
C_{1}\\D_{1}
\end{pmatrix}
=\begin{pmatrix}
1\\0
\end{pmatrix}
\end{equation}
\begin{equation}\label{eq:system2}
\begin{pmatrix}
M(a,1,w_{1})&e^{w_{1}} U(1-a,1,-w_{1})\\M'(a,1,w_{1})& (e^{w} U(1-a,1,-w))'_{w_{1}}
\end{pmatrix}
\begin{pmatrix}
C_{2}\\D_{2}
\end{pmatrix}
=\begin{pmatrix}
0\\1
\end{pmatrix}
\end{equation}
Solving the above two systems and using the relations (see Abramowitz~\cite{Abramowitz:1964}, Chapter 13)
\begin{equation}
\label{relspecialfunction}
\left.
\begin{array}{rcl}
M'(a,1,w)    &=&     a M(a+1,1+1,w),\\
U'(1-a,1,-w) &=&- (1-a)U(1+(1-a),1+1,-w)(-1).
\end{array}
\right\}
\end{equation}
we obtain
\begin{equation}
\label{coef2}
\left.
\begin{array}{rcl}
C_{1}&=&\displaystyle{e^{w_{1}}\left(U\left(\frac{1-k\lambda}{2},1,-w_{1}\right)+ \frac{1-k\lambda}{2}U\left( \frac{3-k\lambda}{2},2,-w_{1}\right)\right)/ W\lbrace  1,2 \rbrace}\\
D_{1}   &=&-\displaystyle{\frac{1+k\lambda}{2}M\left(  \frac{3+k\lambda}{2},2,w_{1}   \right)/ W\lbrace  1,2 \rbrace}\\
C_{2}   &=&-\displaystyle{ e^{w_{1}} U\left( \frac{1-k\lambda}{2},1,-w_{1}   \right)/ W\lbrace  1,2 \rbrace}\\
D_{2}   &=&\displaystyle{ M\left(  \frac{1+k\lambda}{2},1,w_{1}   \right)/ W\lbrace  1,2 \rbrace}
\end{array}
\right\}
\end{equation}
where $W\lbrace  1,2 \rbrace$ is the determinant of the coefficient matrix of the system \eqref{eq:system1}.

Similar to the procedure of the previous section~\ref{subsection:constant-viscosity}, we find $A$ and $B$ so that $f(-L;\lambda)=\mu_{0}(-L)$ and $f'(-L;\lambda)=\alpha_{1}(k)\lambda +\alpha_{2}(k)$. Therefore, it follows from \eqref{eq:fxi} and \eqref{eq:cond2} that
\begin{equation*}
\begin{array}{rcl}
A e^{-kL}                &=&\mu_{0}(-L),\\
Ake^{-kL}-2kB e^{-kL}&=&\alpha_{1}(k)\lambda +\alpha_{2}(k),
\end{array}
\end{equation*}
and therefore $A=\mu_{0}(-L)e^{kL}$ and $B=-\left(\frac{\alpha_{1}(k)\lambda+\alpha_{2}(k)-k\mu_{0}(-L)}{2k}\right)e^{kL}$. Substituting these constants in the function $f(\xi;\lambda)$ defined by \eqref{eq:fxi}, we obtain a solution of the ODE that satisfies the boundary condition at $\xi=-L$ of the eigenvalue problem \eqref{eq:main-evp}. Since $A$ and $B$ depend on the spectral parameter $\lambda$, it follows that the eigenvalues of the problem \eqref{eq:main-evp} can be obtained by studying the following algebraic equation which is a reformulation of the boundary condition at $\xi=0$ of the eigenvalue problem \eqref{eq:main-evp}.
\begin{equation}\label{al2}
\mu_{0}(0)\frac{f_\xi(0;\lambda)}{f(0;\lambda)}=\beta_{1}(k)\lambda -\beta_{2}(k).
\end{equation}
Since the right-hand side of the above equation does not depend on $L$, we need to study the asymptotic limits ($L\to 0$ and $L\to\infty$) of the lefthand side of (\ref{al2}). Notice that 
the expression $\displaystyle{{f_{\xi}(0;\lambda)}/{f(0;\lambda)}}$ above is given by (see \eqref{eq:fxi})
\begin{equation}\label{eq:ratio1}
\frac{f_{\xi}(0;\lambda)}{f(0;\lambda)}= k\left(1-2\frac{Az'_{1}(w_{2})+Bz'_{2}(w_{2})}{Az_{1}(w_{2})+Bz_{2}(w_{2})}\right).
\end{equation}
Therefore, we first find the asymptotic approximations for $z_1(w_2), z_2(w_2), z_1'(w_2)$, and $z_2'(w_2)$ in both cases below before estimating the ratio ${f_{\xi}(0;\lambda)}/{f(0;\lambda)}$ using \eqref{eq:ratio1} for its use in \eqref{al2}. Below, we write $w_{1}=d_{1}L$ and $w_{2}=d_{2}L$ where $d_1$ and $d_2$ are given by (see \eqref{eq:w1-w2-def}),
\begin{equation}\label{factores}
d_{1}=-2k\mu_{0}(-L)/(\mu_{0}(0)-\mu_{0}(-L)),\quad \textrm{and}\quad d_{2}= -2k(\mu_{0}(0))/(\mu_{0}(0)-\mu_{0}(-L)).
\end{equation}
\vskip 0.1truein

\noindent\underline{\bf First case} (When $L\to\infty$):\ It follows from Abramowitz and Stegun~\cite{Abramowitz:1964} that
\begin{equation}\label{asymphypergeometric1}
\left.
\begin{array}{rcl}
M(a,1,d_{j}L)&=&\displaystyle{\frac{\Gamma(1)}{\Gamma(a)}(-d_{j}L)^{-a}(1+O(|L|^{-1}))},\quad
\textrm{as}\ L\to\infty,\ \ \textrm{for}\ \ j=1,2\\
U(1-a,1,-d_{j}L)&=&(-d_{j}L)^{-(1-a)}(1+O(|L|^{-1})),\quad \textrm{as}\ L\to\infty,\ \ \textrm{for}\ \ j=1,2.
\end{array}
\right\}
\end{equation}
Using the identities from (\ref{relspecialfunction}) we obtain
\begin{equation}
\label{asymphypergeometric2}
\left.
\begin{array}{rcl}
M'(a,1,d_{j}L)&=&\displaystyle{\frac{\Gamma(1+1)}{\Gamma(1+a)}(-d_{j}L)^{-(1+a)}(1+O(|L|^{-1}))},\quad
\textrm{as}\ L\to\infty,\ \ \textrm{for}\ \ j=1,2\\
U'(1-a,1,-d_{j}L)&=&(1-a)(-d_{j}L)^{-(1+(1-a))}(1+O(|L|^{-1})),\quad
\textrm{as}\ L\to\infty,\ \ \textrm{for}\ \ j=1,2.
\end{array}
\right\}
\end{equation}
Using \eqref{coef2}, \eqref{asymphypergeometric1} and the relation $\displaystyle{\lim_{L\to\infty}{e^{d_{2}L}}/{e^{d_{1}L}}=0}$ in the expression $\eqref{eq:two-lin}_1$ for $z_1(w_2)$, we obtain
$$z_{1}(w_{2})=O\left\lbrace e^{d_{1}L}(-d_{1}L)^{-(3-k\lambda)/2} (-d_{2}L)^{-(1+k\lambda)/2}\right\rbrace $$
which can be written as $z_{1}(w_{2}) \sim C_{1}M(a,1,w_{2}),$ where
$\displaystyle{C_{1}=O\lbrace e^{d_{1}L}(-d_{1}L)^{(-3+k\lambda)/2} \rbrace}$ (see \eqref{coef2}). Using similar arguments it follows that
\begin{equation}
\label{asymptotics}
\left.   \begin{array}{rcl}
                 z_{1}(w_{2}) &\sim& C_{1}M(a,1,d_{2}L)\\
                 z_{2}(w_{2}) &\sim& C_{2}M(a,1,d_{2}L)\\
                 z'_{1}(w_{2}) &\sim& C_{1}M'(a,1,d_{2}L)=C_{1}aM(a+1,1+1,d_{2}L)\\
                 z'_{2}(w_{2}) &\sim& C_{2}M'(a,1,d_{2}L)=C_{2}aM(a+1,1+1,d_{2}L)
                 \end{array}\right\}
\end{equation}
for $L\to\infty$, see (\ref{coef2}) for the dependence of  $\lambda,\ d_{1}$ and $L$ of the coefficient $C_{1},\ D_{1},\ C_{2}$ and $D_{2}$. Thus, using the above asymptotic results for the coefficient $C_{1},\ D_{1},\ C_{2}$ and $D_{2}$ and the asymptotic results for the confluent hypergeometric functions given in (\ref{asymphypergeometric1}) and (\ref{asymphypergeometric2}), we get
\begin{equation}
\label{cociente}
\left.
\begin{array}{rcl}
\displaystyle{\lim_{L\to\infty}}\frac{Az'_{1}(w_{2})+Bz'_{2}(w_{2})}{Az_{1}(w_{2})+Bz_{2}(w_{2})}&=&\left(\frac{1+k\lambda}{2}\right)
\displaystyle{\lim_{L\to\infty}}\frac{M(1+a,2,d_{2}L)(C_{1}A+C_{2}B)}{M(a,1,d_{2}L)(C_{1}A+C_{2}B)}\\
&=&\left(\frac{1+k\lambda}{2}\right)\displaystyle{\lim_{L\to\infty}\frac{\frac{\Gamma(2)}{\Gamma(1+a)}(-d_{2}L)^{-(1+a)}}
{\frac{\Gamma(1)}{\Gamma(a)}}(-d_{2}L)^{-a}}\\
&=&0.
\end{array}
\right\}
\end{equation}
Substituting this in \eqref{eq:ratio1}, we obtain
\begin{equation*}
\lim_{L\to\infty} \frac{f_{\xi}(0;\lambda)}{f(0;\lambda)} = k.
\end{equation*}
Therefore, equation (\ref{al2}) becomes $\mu_{0}(0)k=\beta_{1}(k)\lambda -\beta_{2}(k)$. Using $\sigma=U_{0}/\lambda$ and expressions for the coefficients from $\eqref{coef}_2$, we obtain
$$\sigma=kU_{0}\frac{(\mu_{2}-\mu_{0}(0))}{(\mu_{2}+\mu_{0}(0))}-\frac{Tk^{3}}{(\mu_{2}+\mu_{0}(0)).}$$
which is the classical formula for Saffman-Taylor instability of the leading interface. Similarly, we can also recover the the classical formula for Saffman-Taylor instability of the trailing interface by reversing the shooting technique as discussed at the end of section \ref{subsection:constant-viscosity}.
\vskip0.1truein

\noindent\underline{\bf Second case} (When $L\to 0$):\ Similar to the previous case, we will first need to get asymptotic approximations  for $z_1(w_2), z_2(w_2), z_1'(w_2)$, and $z_2'(w_2
)$ in this limit. Notice that in this case, singularities of the confluent hypergeometric function of the second kind will arise. Now, we give the following asymptotic results from Abramowit
z and Stegun~\cite{Abramowitz:1964}
\begin{equation}
\label{asymphypergeometric3}
\left. \begin{array}{rcl}
               U(1-a,1,-d_{j}L)&=& \displaystyle{- \frac{1}{\Gamma(1-a)}\left( \ln(\vert d_{j}L  \vert)  +\Psi(1-a)  \right)   +O(L\ln L)}\\
               U(2-a,2,-d_{j}L)&=& \displaystyle{\frac{\Gamma(2-1)}{\Gamma(2-a)} \vert d_{j} L \vert^{1-2} +O(\ln L),      }
               \end{array}
\right\}
\end{equation}
where we recall that $d_{1}$ and $d_{2}$ are defined by \eqref{factores}. Similar to the calculations of the previous case $L\to\infty$, we present the dominant terms of the left hand side of (\ref{al2}). It is worth pointing out that due to \eqref{asymphypergeometric3}, the derivative  of the confluent hypergeometric function of the second kind is dominant.
From the definition of the coefficients $C_{1},\ D_{1},\ C_{2}$ and $D_{2}$ given in \eqref{coef2} and the asymptotic results presented in \eqref{asymphypergeometric3}, we obtain
\begin{equation}
\label{asymz1}
\begin{array}{rcl}
z_{1}(d_{2}L)        &\sim& \displaystyle{\frac{1}{W\lbrace  1,2\rbrace} \left( \frac{1-k\lambda}{2}  \right)\left\lbrace  \frac{\Gamma(1)}{\Gamma(1-a)}\right\rbrace\vert d_{1}L\vert^{-1}}\\
z'_{1}(d_{2}L)&\sim&  \frac{1}{W\lbrace 1,2 \rbrace}\left(\frac{1}{2}+\frac{k\lambda}{2}\right)\left(\frac{1}{2}-\frac{k\lambda}{2}\right)\displaystyle{\frac{\left\lbrace \frac{1}{\vert d_{1}L\vert}-  \frac{1}{\vert d_{2}L\vert} \right\rbrace}{\Gamma(1+a)}}.\\
\end{array}
\end{equation}
We remark that
\begin{equation*}
z'_{1}(d_{2}L)\sim C_{1}\left(\frac{1+k\lambda}{2}\right)+D_{1}\left(\frac{1-k\lambda}{2}\right)\frac{\Gamma(1)}{\Gamma(2-a)}|d_{2}L|^{-1}
\end{equation*}
and therefore the asymptotic result for $z'_{1}(w_{2})$ follows from the definition of the coefficient $C_{1}$ and $D_{1}$, see (\ref{coef2}). From the forms of $C_{1},\ D_{1},\ C_{2}$ and $D_{2}$, we get $z_{2}(w_{2})=o(z_{1}(w_{2})).$ Therefore
\begin{equation}
\label{asymz2}
\frac{Az_{1}(w_{2}) + Bz_{2}(w_{2})}{Az_{1}(w_{2})}\sim 1.
\end{equation}
Similarly, we obtain
\begin{equation}
\label{asymz'2}
\left.
\begin{array}{rcl}
z_{2}'(w_{2})& \sim & -D_{2}e^{w_{2}}U'(1-a,1,-w_{2})\\
             & \sim & \displaystyle{\frac{1}{W\lbrace 1,2 \rbrace}\left(\frac{1-k\lambda}{2}\right)U(2-a,2,-w_{2})e^{w_{2}}}\\
             & \sim &  \displaystyle{\frac{1}{W\lbrace 1,2 \rbrace}\left(\frac{1-k\lambda}{2}\right)\left\lbrace \displaystyle{ \frac{\Gamma(1)}{\Gamma(2-a)}} \right\rbrace|d_{2}L|^{-1}}\\
             &\sim&\frac{1}{\Gamma(1-a)}\frac{1}{W\lbrace 1,2\rbrace}\frac{1}{|d_{2}L|}
\end{array}
\right\}
\end{equation}
Using \eqref{asymz1}, \eqref{asymz2}, and \eqref{asymz'2}, it follows that
\begin{equation*}
\begin{array}{rcl}
\displaystyle{\lim_{L\to 0}}\frac{f_{\xi}(0;\lambda)}{f(0;\lambda)}&=& k\left(1-2\displaystyle{\lim_{L\to0}\frac{Az'_{1}(w_{2})+Bz'_{2}(w_{2})}{Az_{1}(w_{2})+Bz_{2}(w_{2})}}\right)\\
&=&k\left(1-2\displaystyle{ \lim_{L\to 0}\frac{Az'_{1}(w_{2})+Bz'_{2}(w_{2})}{Az_{1}(w_{2})}}\right)\\
\\
&=&k\left(1-2\displaystyle{ \lim_{L\to 0}\frac{\frac{Aa}{\Gamma(1-a)}\left( \frac{1}{|d_{1}|} - \frac{1}{|d_{2}|} \right)\frac{1}{L}+B\left(  \frac{1}{\Gamma(1-a)}  \right)\left( \frac{1}{|d_{2}L|}  \right) }{A\left( \frac{1}{\Gamma(1-a)}  \right)\left( \frac{1}{|d_{1}L|} \right)}}\right)\\
\\
&=&k\left(1-2\displaystyle{\lim_{L\to 0} \frac{Aa\left( \frac{1}{|d_{1}|} - \frac{1}{|d_{2}|} \right)|d_{1}| + B\frac{|d_{1}|}{|d_{2}|}}{A}}\right)\\
\\
&=&k\left(1-2\displaystyle{ \lim_{L\to 0} \frac{Aa\left( 1 - \frac{|d_{1}|}{|d_{2}|} \right) + B\frac{|d_{1}|}{|d_{2}|}}{A}}\right).
\end{array}
\end{equation*}
From the definition of $d_{1}$ and $d_{2}$ given in \eqref{factores} we obtain $d_{1}/d_{2}=\mu_{0}(-L)/\mu_{0}(0)$ and therefore
\begin{equation*}
\lim_{L\to 0}\frac{f_{\xi}(0;\lambda)}{f(0;\lambda)}=k\left[1-2\left(\frac{1+k\lambda}{2}\right)\left(\frac{\mu_{0}(0)-\mu_{0}(-L)}{\mu_{0}(0)}  \right)-2\frac{B}{A}\frac{\mu_{0}(-L)}{\mu_{0}(0)}\right],
\end{equation*}
where $A=\mu_{0}(-L)e^{kL}$ and $B=-\left(\frac{\alpha_{1}(k)\lambda+\alpha_{2}(k)-k\mu_{0}(-L)}{2k}\right)e^{kL}$. It then follows that
\begin{equation*}
\lim_{L\to 0} \mu_{0}(0)\frac{f_{\xi}(0;\lambda)}{f(0;\lambda)} = \left[ \alpha_{1}(k)-k^2(\mu_{0}(0)-\mu_{0}(-L))\right]\lambda+\alpha_{2}(k).
\end{equation*}
Using this in equation (\ref{al2}), we obtain
$$\left[ \alpha_{1}(k)-k^2(\mu_{0}(0)-\mu_{0}(-L))\right]\lambda+\alpha_{2}(k)=\beta_{1}(k)\lambda-\beta_{2}(k)  $$
which is equivalent to
$$\left((\alpha_{1}(k)-\beta_{1}(k))-{k^2}(\mu_{0}(0)-\mu_{0}(-L))\right)\lambda=-\alpha_{2}(k)-\beta_{2}(k).$$
After substituting the values of $\alpha_1(k)$, $\alpha_2(k)$, $\beta_1(k)$ and $\beta_2(k)$ and simplifying we obtain
\begin{equation*}
\begin{array}{c}
\left(\frac{S+T}{U_{0}}k^{4}+k^{2}(\mu_{1}-\mu_{2}\right)\lambda=-k(\mu_{2}+\mu_{1}).
\end{array}
\end{equation*}
Therefore,
$$\sigma=-\frac{(S+T)}{(\mu_{2}+\mu_{1})}k^{3}+U_{0}k\frac{(\mu_{2}-\mu_{1})}{(\mu_{2}+\mu_{1})},$$
which is the formula for the growth rate of an interface with surface tension $(S+T)$, which is what should be expected in this limit. Thus we recover the classical formula for the growth rate in this limit.

\section{Conclusions}\label{conclusions}
We converted a non-standard eigenvalue problem arising in the linear stability analysis of a three-layer Hele-Shaw model of enhanced oil recovery to a boundary value problem for 
Kummer's equation when the middle layer has a linear viscous profile. We presented the general solution in terms of Frobenius series and discussed the convergence properties of these 
series solutions. We also formally gave the dispersion relation implicitly through the existence criterion for non-trivial solutions.
In order to recover the well-known physical solutions for some limiting cases, we rewrote the general solutions using a different set of fundamental solutions and analyzed these 
for those limiting cases: (i) when the viscous profile of the middle layer approaches a constant viscosity, both in the case of a fixed-length middle layer and also as the length 
of the middle layer appraoches infinity; and (ii) when the length of the middle layer approaches zero. We showed that we were thus able to recover the correct physical solutions.

\medskip

\section*{\bf Acknowledgments:}\
This paper was made possible by an NPRP grant \# 08-777-1-141 to one of the authors (Prabir Daripa) from the Qatar National Research Fund (a member of the Qatar Foundation). The second author (Oscar Orellana) acknowledges financial support through this grant for travel to TAMUQ, Qatar for a two day workshop on ``International Workshop on Enhanced Oil Recovery and Porous Media Flows" organized by the first author (Prabir Daripa) during July 31st and August 1 of 2013. The work of the second author (Oscar Orellana) was also supported in part by Fondo Nacional de Desarrollo Centifico y Technologico (FONDECYT) under grant 1141260 and Universidad Tecnica Federico Santa Maria, Valparaiso, Chile.
The statements made herein are solely the responsibility of the authors.
\bigskip

\section*{Appendix: Kummer's Equation}\label{section:appendix3}

Kummer's equation has the general form
\[
 w \frac{d^2z}{dw^2} + (b-w) \frac{dz}{dw} -  az = 0,
\]
where $b=1$ and $a=\frac12(1+\frac{kU_0}\sigma).$ The two linearly independent solutions are $z_1(a,b,w)$ and $z_2(a,b,w)$ where the general expression for $z_1(a,b,w)$ is given by
\begin{equation*}
z_1(a,b,w) = \frac{a_0}{b_0}+\frac{a_1w}{b_1} + \frac{a_2w^2}{b_2\,2!} +
\frac{a_3w^3}{b_3\,3!} +\cdots+ \frac{a_nw^n}{b_n\,n!} +\cdots
\end{equation*}
where
\begin{align*}
&a_0=1, a_1=a, a_n=a(a+1)(a+2)\cdots (a+n-1),\quad{\rm for}\ n=2,3,\cdots\\
&b_0=1, b_1=b, b_n=b(b+1)(b+2)\cdots (b+n-1),\quad{\rm for}\ n=2,3,\cdots
\end{align*}
The linearly independent second solution $z_2(a,b,w)$ is similarly given by a series which can be easily constructed by the method of Frobenius.

\end{document}